# Determination of the paradihlorobenzene and paradibromobenzene solid solutions nanoparticles structure via Raman spectra

M.A. Korshunov

L.V. Kirensky Institute of Physics Siberian Branch of RAS, 660036 Krasnoyarsk, Russia

*e-mail:* kors@iph.krasn.ru

**Abstract.** We measured the small frequencies Raman spectrum of the paradihlorobenzene and paradihlorobenzene solid solution nanoparticles with the size ~100 nanometers. Values of frequencies of lines decrease. The size of nanoparticles was determined by the electronic microscope. Calculations of nanoparticles structure were done using the method of molecular dynamics and histograms of nanoparticles spectra were calculated via the Dyne's method. The result is that the Raman spectrum is the sum of spectra from the central part of the nanoparticle and superficial structures with smaller concentration of paradihlorobenzene.

Displacement of frequencies of lines in low-frequency area it is marked in work [1] and in a spectrum organic molecular nano particle [2]. But as the spectrum with reduction of the size of particles for such crystals doped an impurity not absolutely clearly will change. The spectrum can change depending on that as the impurity in one nano particles is distributed.

For studying of this question in the present work comparison of experimental spectra of combinational dispersion of light lattice fluctuations nano particles firm solutions paradihlorobenzene with paradibromobenzene with settlement spectra nano particles has been spent at change of their size and concentration change in nano particles. For spectrum interpretation lattice fluctuations calculations on a method the dyne [3] are carried out. The structure nano particles was modelled by means of a method of molecular dynamics [4]. Massive monocrystals of firm solutions paradihlorobenzene with paradibrombenzene are well enough studied by various methods. There is an interpretation lattice fluctuations of this monocrystal [5,6].

For reception organic nano particles often use crushing of initial substance [7]. The second used method is a dusting on integumentary glass of a studied firm solution for evaporation reduction the received sample was covered from above with other integumentary glass. In both methods at evaporation the size of particles that allows to receive particles of the necessary size decreases. The sizes nano particles were defined on an electronic microscope. After that record of spectra of combinational dispersion of light nano particles on spectrometer **jobin yvon** 64000 was spent. The received spectra of combinational dispersion of light of small frequencies nano particles ~100nm. The size of concentration of an impurity was defined on relative intensity of valency intramolecular fluctuations. In drawing 1 the spectrum lattice fluctuations of a firm solution paradihlorobenzene with paradibromobenzene is presented at the size of particles of ~100 nanometers (a), result of division of lines (b).

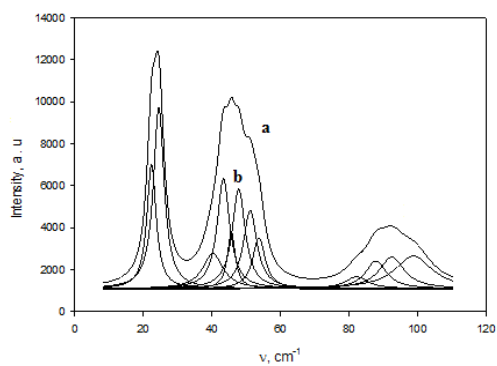

**Drawing 1.** A spectrum lattice fluctuations of a firm solution at the size of particles of ~100 nanometers (a), result of division of lines (b).

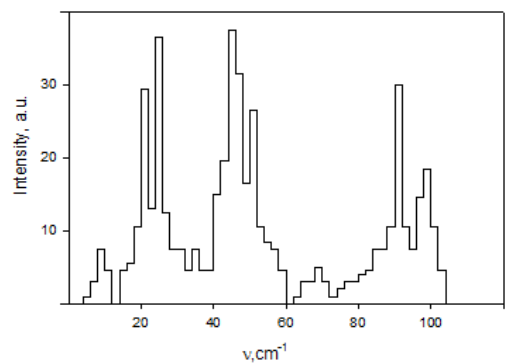

**Drawing 2.** Histograms of frequencies lattice fluctuations nano particles with a size ~ 100 nanometers with structure when concentration of molecules paradihlorobenzene decreases to particle border.

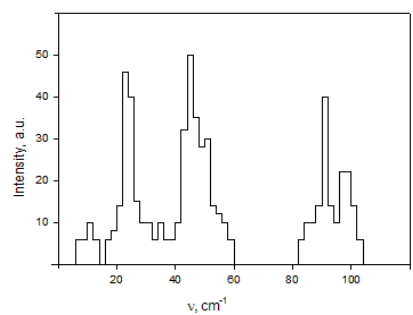

**Drawing 3.** Histograms of frequencies lattice fluctuations nano particles with a size ~ 100 nanometers for a mix nano particles, each of which has concentration of components or as in the centre nano particles from the previous calculation or on border.

Using a method of molecular dynamics and calculation of spectra lattice fluctuations it is possible to study structure and formation processes nano particles at level of movement of separate molecules.

interaction, received by us [9] were used earlier at which settlement spectrum lattice fluctuations of monocrystals paradihlorobenzene, paradibromobenzene and their firm solutions coincided with values of frequencies of lines of experimental spectra lattice fluctuations of the studied monocrystals received at polarising researches.

at first the structure of a single-layered molecular film of a firm solution paid off at the set concentration of components with increase in quantity of layers this film forms nano particles using the found structure, calculations of spectra lattice fluctuations on a method the dyne have been carried out.

The calculated histograms of spectra are presented in drawing 3. Histograms for two cases have been calculated. The spectrum from one nano particles in which concentration of components changes from the centre to border (drawing 2) or consists of a set of spectra of separate particles with different concentration (drawing 3).

It is visible that the spectrum from one nano particles in which concentration of components changes from the centre to border is close to an experimental spectrum.

The structure of molecules was accepted absolutely rigid. The interaction potential has been chosen in shape (6-exp) [8]. In calculations factors in the potential of the

Apparently from drawing 2 in a spectrum it is observed twelve intensive lines. Six from which belong to a spectrum lattice fluctuations paradihlorobenzene with paradibromobenzene at concentration of components of 86 % paradihlorobenzene, and the second set of lines, probably, concerns a spectrum of a firm solution with concentration paradihlorobenzene 70 %.

The carried out calculations on a method of molecular dynamics of structure nano particles a firm solution paradihlorbenzene with paradibrombenzene have shown that concentration of molecules paradihlorobenzene in the centre nano particles is more than at its borders (on visible this distribution depends on temperature). It is connected by that communication of molecules paradihlorobenzene with the next molecules is less than similar interaction of molecules paradibromobenzene. Therefore at molecule evaporation paradihlorobenzene on border with greater probability disappear increasing concentration paradibromobenzene. It can find reflexion in a spectrum nano particles a spectrum from the central part with It is more concentration paradihlorobenzene has in a spectrum of frequency of lines above, than a spectrum from superficial area with smaller concentration paradihlorobenzene . In a spectrum of a similar particle should be observed not only fluctuations from a volume part, but also and superficial fluctuations.

At calculations reduction of frequencies of lines is found. It is caused by increase in parametres of a lattice in nano particles in comparison with a volume crystal, as observed in experimental spectra.

The comparative researches of change of spectra conducted in given work nano particles firm solutions at reduction of their sizes and calculations of their structure by a method of molecular dynamics, and also calculation of spectra lattice fluctuations supply the additional information on structure and dynamics of a lattice. At modelling of structure of particles on a method of molecular dynamics and calculation of histograms of spectra lattice fluctuations on a method the dyne, it is found that at reduction of the sizes of particles lattice parametres increase, but non-uniformly both in directions, and on diameter nano particles, concentration of molecules paradihlorobenzene in structure nano particles in the central part more and decreases to edge. It causes the doubling of a spectrum observed in experimental spectra lattice of fluctuations.

## References

[1] **C.C.YANG AND S.LI**. J. Phys. Chem., **112b** (2008 14193-14197

[2] **M.A.KORSHUNOV**. Wholesale. And a spectrum., 109(2010) 1387-1390

[3] **P.V.DIN**. Computing methods in the theory of a firm body. Moscow, the world (1975)

[4] **D.C.RAPAPORT**. The art of molecular dynamics simulation. Cambridge, university Press (1995 553

[5] **V.F.SHABANOV, V.P.SPIRIDONOV, M.A.KORSHUNOV**.J.Apl. spectrum., 25(1976) 698-701

[6] **A.I.KITAJGORODSKY.** Organic Crystallochemistry. Moscow, publishing house. AS of the USSR (1955) 558

[7] **A.J.UTEHINA, G.B.SERGEEV**. Successes of chemistry., 10(2011) 233-248

[8] **A.I.KITAJGORODSKY.** Molecular crystals. Moscow, the science (1971)

[9] **V.F.SHABANOV, M.A.KORSHUNOV.** Sol. St. Phys., 37(1995) 3463-3469